\documentstyle[12pt,epsf]{article}
\newcommand{\beq}{\begin{equation}}
\newcommand{\eeq}{\end{equation}}
\newcommand{\bea}{\begin{eqnarray}}
\newcommand{\eea}{\end{eqnarray}}
\newcommand{\rf}[1]{(\ref{#1})}
\newcommand{\C}[1]{{\cal #1}}
\newcommand{\tmu}{\tilde\mu}
\newcommand{\mubar}{\bar\mu}
\newcommand{\half}{{1\over 2}}
\newcommand{\threehalves}{{3\over 2}}
\newcommand{\third}{{1\over 3}}
\newcommand{\quarter}{{1\over 4}}
\newcommand{\halfd}{{d\over 2}}
\newcommand{\gstr}{\gamma}

\newcommand{\et}{\eta}

\newcommand{\nn}{\nonumber}

\begin{document}
\topmargin 0pt
\oddsidemargin 5mm
\headheight 0pt
\topskip 0mm

\addtolength{\baselineskip}{0.20\baselineskip}

\pagestyle{empty}

\begin{flushright}
OUTP-96-45P\\
5th August 1996\\
hep-th/9608021
\end{flushright}

\begin{center}

\vspace{18pt}
{\Large \bf A Simple Model of Dimensional Collapse}

\vspace{2 truecm}

{\sc Jo\~ao D. Correia\footnote{e-mail: j.correia1@physics.ox.ac.uk} 
and John F. Wheater\footnote{e-mail: j.wheater1@physics.ox.ac.uk}}

\vspace{1 truecm}

{\em Department of Physics, University of Oxford \\
Theoretical Physics,\\
1 Keble Road,\\
 Oxford OX1 3NP, UK\\}

\vspace{3 truecm}

\end{center}

\noindent
{\bf Abstract.} We consider a simple model of $d$ families of scalar 
field interacting with geometry in two dimensions. The geometry is locally flat
and has only global degrees of freedom.  When $d<0$ the universe is locally 
two dimensional but for $d>0$ it collapses to a one dimensional manifold.
The model has some, but not all, of the characteristics believed to be features of the
full theory of conformal matter interacting with quantum gravity which has
 local geometric degrees of freedom.

\vfill
\begin{flushleft}
\end{flushleft}
\newpage
\setcounter{page}{1}
\pagestyle{plain}

It is known that conformal matter interacting with quantum gravity on a two dimensional
manifold of fixed topology alters both the local and global nature of the geometry \cite{BP,FD,KKM}. 
For matter with central charge $c<1$ many of the properties of the system can be calculated
through the KPZ results \cite{KPZ}.  In this regime the system remains locally two 
dimensional but has global properties that vary with $c$. In the formal limit that 
$c\to -\infty$ we recover the semi-classical  behaviour of matter interacting with a flat 
background metric; the KPZ results are non-analytic at $c=1$ and for $c>1$  they do 
not make sense. For very large and positive $c$ the geometry is known to be 
dominated by singular, branched polymer like, configurations of the metric
\cite{BP} but the regime
just above $c=1$ is not completely understood \cite{Bre,Wex,ADJT,mghjfw}.

In this paper we consider some simpler models of matter interacting with geometry in two
dimensions; in these models the geometry is always locally flat and the only geometrical 
degrees of freedom are global parameters. The models show similar behaviour to that described
above but with the crucial difference that the regime analogous to $c>1$ can be analyzed 
exactly. We will show that these systems undergo a phase transition from a small $c$ phase 
in which space is locally two dimensional to a large $c$ phase in which it is locally one 
dimensional.

In order to be locally flat everywhere the space must have the topology of a torus. The models
are defined on a square lattice of lattice spacing $a$ with sides of length $L_{1,2}$ spacings; 
in the 1 direction we
impose periodic boundary conditions while in the 2 direction we impose helical boundary 
conditions, being shifted by $M$ lattice spacings. The matter content consists of $d$ 
scalar fields, $X$, whose partition function on such a lattice we denote by $Z^X(L_1,L_2,M)$.
 We define two versions of the model by the
partition functions
\bea \C Z^I(\mu,d)&=&\sum_{L_1,L_2}e^{-\mu L_1L_2}\left(Z^X(L_1,L_2,0)\right)^d\cr
 \C Z^{II}(\mu,d)&=&\sum_{L_1,L_2}\sum_{M=0}^{L_1-1}e^{-\mu L_1L_2}\left(Z^X(L_1,L_2,M)\right)^d\label{latdef}\eea
where $\mu$ is the cosmological constant. Model I does not sum over twists whereas
model II does and therefore includes all regular squarings of the torus.
In continuum language model I sums over the imaginary part of the modular parameter
 $\tau={M+iL_2\over L_1}$.
 whereas 
model II sums over both real and imaginary part; note however that the measure implied by
\rf{latdef} is not the modular invariant measure.

These partition functions are convergent sums provided that $\mu$ is greater than
some critical value $\mu_c$.
In the usual notation for quantum gravity the string exponent $\gstr$ is defined
by the leading non-analytic behaviour of the partition functions as $\mu\downarrow\mu_c$
\beq \C Z =P(\mu)-Q(\mu)(\mu-\mu_c)^{2-\gstr}\label{gammadef}\eeq
where $P$ and $Q$ are regular functions at $\mu=\mu_c$. In fact for full gravity
with conformal matter of $c<1$ on a torus we expect that $\gstr=2$ which means
that there is a logarithmic singularity in $\C Z$.

The sums arising in $\C Z^{I,II}$ at $d=0$ are associated with Jacobi functions
\cite{kazakov}.
However elementary methods are sufficient to elucidate the critical properties.
Firstly it is clear that, because the number of graphs only increases polynomially
in system size, the partition functions converge for $\mu>0$ so that $\mu_c=0$.
Now consider
\beq\C Z^I=\sum_{L_1,L_2}e^{-\mu L_1 L_2}=\sum_L {1\over e^{\mu L}-1}\eeq
The summand is monotonic decreasing so
\beq \C Z^I={1\over e^{\mu }-1}+\int_2^\infty {dL\over e^{\mu L}-1}+\C R\eeq
where
\bea |\C R|&\le&\sum_{L=2}^\infty\left\{{1\over e^{\mu (L-1)}-1}-{1\over e^{\mu L}-1}
\right\}\nn\\
&\sim& {C\over \mu}+O(1)\eea
with $C$ a constant. The integral is elementary and we find that,
as $\mu\to 0$,
\beq\C Z^I=-{\log\mu\over\mu}+{C'\over \mu}+O(1)\label{logdiv}\eeq
Comparing \rf{logdiv} with \rf{gammadef} we see that $\gamma^I=3$
but with a logarithmic correction.

It is informative to look at a slightly modified version of the $d=0$ model
\beq \widetilde\C Z^I=\sum_{L_1}\sum_{L_2=1}^{L_1} e^{-\mu L_1L_2-\zeta{L_1\over L_2}}\eeq
in which tori with $t\ne 1$ are suppressed by an amount depending on $\zeta \ge 0$. Now
\bea \widetilde\C Z^I&=&\sum_{L=1}^\infty{ e^{-(L-1)(\mu L+{\zeta\over L})}\over
 e^{\mu L+{\zeta\over L}}-1}\nn\\
&=&\sum_{L=1}^\infty{ e^{-(L-1)(\mu L+{\zeta\over L})}\over \mu L+{\zeta\over L}}
+\C R'\eea
where, on account of $\vert(e^x-1)^{-1}-x^{-1}\vert < \half$,
\bea \vert\C R'\vert &<& \half \sum_{L=1}^\infty e^{-(L-1)(\mu L+{\zeta\over L})}\nn\\
   &\sim& {1\over \sqrt\mu}\eea
The remaining sum can be majorized by dropping $\mu$ in the denominator
leaving, for non-zero $\zeta$, as $\mu\to 0$
\beq \widetilde\C Z^I < {1\over\zeta\mu}+{C''\over\sqrt\mu}\label{bigz}\eeq
Thus we see that for arbitrary $\zeta > 0$ the logarithmic divergence in
\rf{logdiv} is removed.  In fact for all finite non-zero $\zeta$ the divergence
 is $\mu^{-1}$
 and at large $\zeta$ it is easy to see that \rf{bigz} is an equality; only when
 $\zeta$ is strictly infinite do we recover the $\mu^{-\half}$ behaviour expected for
a single torus with $L_1=L_2$\footnote{This phenomenon is, at least superficially,
 similar to that recently
demonstrated for $R^2$ gravity \cite{RR}; for all finite values of the $R^2$ coupling
the system remains in the pure gravity universality class
and only at infinite coupling does it become flat.}

Similar considerations  for model II lead to the results
\beq \C Z^{II} \sim {1\over \mu^2}+\ldots\eeq
and, for $\zeta >0$,
\beq \widetilde\C Z^{II} \sim {1\over \mu^{{3\over 2}}}+\ldots\eeq
Thus model II has $\gamma^{II}=4$ at $\zeta=0$ \cite{kazakov} and $\gamma^{II}={7\over 2}$ for
$\zeta>0$.

A change in the value of $\gamma$ indicates a phase transition in which the 
nature of the geometry changes. The best known example of this is probably 
the Ising model interacting with quantum gravity in a two dimensional
universe of spherical topology.  For
all values of the Ising coupling except the critical value we find
 $\gstr=-\half$;
but at the critical coupling, where the Ising correlation functions are 
long range and can affect the global geometry, $\gstr=-\third$ \cite{kk}.

 Adding matter
fields to the model introduces an exponential dependence on the modular parameter of 
similar form to that in $\widetilde\C Z^{I,II}$.
To calculate the partition functions when $d\ne 0$  we will first use the
continuum formulation.  The area $A$ 
of the torus is given by  $L_1 L_2a^2$  and the modular
parameter $\tau=s+it={M+iL_2\over L_1}$;  note that $A\ge a^2t$.
The partition function of a single gaussian field is then \cite{ItZu}
\beq Z^X=e^{f_0 A}t^{-\half} {1\over \vert\eta(\tau)\vert^2}\label{ItZu}\eeq
where $f_0$ is some constant with dimensions of inverse area
and the Dedekind $\et$ function is given by
\bea \et(\tau)&=&q^{1\over 24} P(\tau),\quad  P(\tau)=\prod_{n=1}^\infty(1-q^n),\quad 
q=e^{i2\pi\tau}\eea
In the region $t\ge 1$, $0\le s <1$  we have the  the obvious inequality
\beq \prod_{n=1}^\infty(1-e^{-2n\pi t})^2 \le \vert P\vert^2\le 
\prod_{n=1}^\infty(1+e^{-2n\pi t})^2\eeq
and elementary arguments yield the bounds
\beq \exp\left({2\log(1-e^{-2\pi t})\over1-e^{-2\pi t}}\right)  \le \vert P\vert^2\le 
\exp\left({2e^{-2\pi t}\over1-e^{-2\pi t}}\right)\label{bound}\eeq
 As we shall see any singularity arising from the integration 
over $t$  arises from the region $t\to\infty$   and therefore
 cannot be affected by the factor
$\vert P\vert^2$ which converges uniformly to 1.

The partition function
$\C Z^{I}$ is then given by
\beq \C Z^{I}(\mu,d)=\int_1^\infty
{dt\over t^{1+\halfd}}\int_t^\infty\,du\,e^{-(\mu-\mu_0d)u}\,
e^{{\pi td\over 6}}\,\vert P(it)\vert^{-2d}\label{ZI}\eeq
 Doing the $u$ integral we get
\beq \C Z^{I}(\mu,d)={1\over \mu-d\mu_0}\int_1^\infty
{dt\over t^{1+\halfd}}\,e^{-t\tmu}\,\vert P(it)\vert^{-2d}\label{ZIa}\eeq
where
\beq \tmu=(\mu-\mu_0d)-{\pi d\over 6}\eeq
Now we can  distinguish three cases
\begin{itemize}
\item  If $d< 0$ then at $\mu=\mu_c=\mu_0d$, $\tmu=-{\pi d\over 6} >0$ and so the
remaining integrals
are finite and
\beq  \C Z^{I}(\mu,d)={C\over \mu-\mu_0d} +\ldots \eeq
as in \rf{bigz}.
\item If $d=0$ then we recover the results \rf{logdiv} that as $\mu\to 0$
\beq \C Z^I(\mu,0)=-{\log\mu\over\mu}+\ldots\eeq
\item If $d>0$ 
the physical singularity is now at $\tmu=0$ since $\tmu< \mu-\mu_0d$ when
$d>0$. Using \rf{bound} to drop the $\vert P\vert^2$ factor and
differentiating  \rf{ZIa}  $1+\lfloor{{d\over 2}}\rfloor$ 
times  with respect to $\mu$
 we obtain a quantity which diverges as
$\tmu\to 0$ and hence
\beq \C Z^I(\mu,d)\sim \tmu^{\halfd}+\C R_1(\mu,d)\label{T3}\eeq
where $\C R_1$ is an analytic function at $\tmu=0$;
 when $d=2m$ ($m$ a positive integer) the 
leading non-analytic behaviour is given by
\beq \C Z^I(\mu,d)\sim \tmu^{\halfd}\log\tmu+\C R_2(\mu,d)\label{T3a}\eeq
\end{itemize}

Similar results can be obtained for model II
\bea  \C Z^{II}(\mu,d<0)&=&{B\over (\mu-d\mu_0)^\threehalves} +\ldots\\
\C Z^{II}(\mu,d=0)&=&{B'\over \mu^2} +\ldots\\
\C Z^{II}(\mu,d>0)&=&B''\tmu^{\halfd-1} +\ldots\label{ZIIc}
 \eea
where $B$ etc are constants. Again, \rf{ZIIc} is modified by a factor of $\log\tmu$ if $d$ is a
positive even integer.

We see that when  $d>0$ the physical singularity comes from the large $t$ 
region  so long thin tori dominate in $\C Z^{I,II}$ and the 
system is essentially one dimensional. On the other hand, when $d<0$ tori with $t\simeq1$ 
dominate and the system is two dimensional. As $d$ increases through zero
there is a phase transition in which
the local  geometrical nature of the system changes from being two dimensional 
to being one dimensional; the value of $\gstr$ also  changes discontinuously
at this point.  There are two competing singularities in $\C Z^{I,II}$; one
is at
\beq \mu_c^{(-)}=\mu_0d\eeq
and the other at 
\beq \mu_c^{(+)}=\mu_0d+{\pi d\over 6}\eeq
Precisely at the point where they coincide, $d=0$, we get a phase transition 
which is first order, the first
derivative of $\mu_c$ with respect to $d$ being discontinuous.

We should be a little careful about drawing conclusions from the continuum
 calculation  in the  $d>0$ regime because the 
most important tori are those near the cut-off (having the length of one
side as large as possible) 
 where 
\rf{ItZu} is no longer necessarily correct.  To check this
  we  examine the lattice version of model I directly.

  The partition function $Z^X$ for a scalar field is
given by
\beq Z^X=\prod_{i=1}^N \int dX_i\;\delta\left(N^{-1}\sum_{k=1}^N X_k\right)
\exp\left(-\half\sum_{<mn>}(X_m-X_n)^2\right)\label{ZN1}\eeq
where $N$ is the number of lattice sites,
 $<mn>$ denotes the link joining neighbouring points labelled by
$m$ and $n$ and the delta function suppresses the zero mode.  The integrals
are easily done yielding
\bea Z^X(L_1,L_2,0)&=&K N^\half e^{-NF}\\
F&=&-\half N^{-1}\sum_{n=0}^{L_1-1}\sum_{m=0}^{L_2-1}
\log\left(\sin^2{m\pi\over  L_2}+\sin^2{n\pi\over  L_1}\right)\label{ZN2}\eea
where the  zero mode is to be excluded and $K$ is a constant.
 We will need $F$ when $L_2\gg L_1$
so first consider the case when $L_1$ is kept finite but $L_2\to \infty$
\bea \mubar(L_1)&=&-{1\over 2\pi L_1}\sum_{n=0}^{L_1-1}\int_0^\pi dx\,
\log\left(\sin^2x +\sin^2{n\pi\over  L_1}\right)\\
&=&\log 2 -{1\over L_1}\sum_{n=0}^{L_1-1}
\log\left(\sin{n\pi\over L_1}+\sqrt{ 1+\sin^2{n\pi\over L_1}}\thinspace\right)   \eea
The finite $L_2$ corrections may be computed using 
the Euler-Maclaurin summation formula; for example  the $n=0$ contribution in the
original sum \rf{ZN2} is given  by 
\bea  S&=&\sum_{m=0}^{L_2-1}
-\half\log\left(\sin^2{m\pi\over  L_2}\right)\nonumber\\
&=&
-{L_2\over 2\pi}\int_{\delta}^{\pi-\delta}\log(\sin^2t)\, dt-\quarter\left\{\log\left(\sin^2\delta\right)+
\log\left(\sin^2(\pi-\delta)\right)\right\}\nonumber\\
& &{}+\delta {B_2\over 2}\Big\{ 
G'(\pi-\delta)-G'(\delta)\Big\}+
\delta^4{B_4\over 4!} \sum_{k=1}^{L_2-2}
G''''\left(\delta(k+\theta)\right)\label{UGH}\eea
where $G(t)=-\half \log(\sin^2t)$, $\delta=\pi/L_1$ and $0<\theta<1$.
It is easy to check that the last two terms of \rf{UGH} vary like $1+O(L_2^{-1})$ and by 
rearranging the integral we find
\beq S=-{L_2\over2\pi}\int_0^\pi \log(\sin^2t)\,dt -\log L_2 +O(1)\eeq
Similar, although more messy, considerations for the $n\ne 0$ case in \rf{ZN2} 
lead to the 
conclusion that
\beq F=\mubar(L_1)
-{\log L_2\over N}+{E\over N}+O(L_2^{-2})\eeq
where $E$ is a constant and hence that 
\beq Z^X(L_1,L_2,0)=K \left({L_1\over L_2}\right)^\half
 e^{N\mubar(L_1)+O(L_1/L_2)}\label{ZN3}\eeq
To compare this with the continuum result the sum  in $\mubar(L_1)$ can be computed
approximately so that  in the regime $L_2\gg L_1\gg 1$,
\bea Z^X&=&K \left({L_1\over L_2}\right)^\half \exp\left\{N\mu_0+{\pi\over 6}{L_2\over L_1} +O(L_1/L_2) +O(L_2/L_1^2)\right\}\nonumber\\
\mu_0&=&\log 2 -{2\C C\over\pi}\eea
where $\C C$ is Catalan's constant in
 agreement with  \rf{ItZu}.

 When  $L_2\gg L_1$ the partition function for a single Gaussian field is given by \rf{ZN3}
so that 
\beq\C Z(\mu,d)=\sum_{L_1=1}^\infty\sum_{L_2=L_1}^\infty e^{-\mu L_1L_2}
\left(e^{-\mubar(L_1) L_1 L_2}\sqrt{L_1\over L_2}\thinspace\right)^d\label{T0}\eeq
Separating out the contribution for $L_1=1$ in \rf{T0} we get
\beq \C Z(\mu,d)=\C F(\mu,d)+\sum_{L_2=1}^\infty L_2^{-{d\over 2}} e^{-L_2(\mu-\mubar(1)d)}\label{T1}\eeq
where
\beq\C F(\mu,d)=\sum_{L_1=2}^\infty\sum_{L_2=L_1}^\infty \left({L_1\over L_2}\right)^\halfd
 e^{- L_1L_2(\mu-\mubar(L_1)d)}\eeq
The second term (corresponding to $L_1=1$) in \rf{T1} diverges at $\mu=\mu_c=\mubar(1)d$;
on the other hand, because $\mubar(x)$ is an increasing function of $x$,
\bea \C F(\mu_c,d)&\le&\sum_{L_1=2}^\infty\sum_{L_2=L_1}^\infty \left({L_1\over L_2}\right)^\halfd
 e^{- L_1L_2d(\mubar(1)-\mubar(2))}\\
&=&\sum_{L_1=2}^\infty\sum_{L_2=L_1}^\infty \left({L_1\over L_2}\right)^\halfd
 e^{- L_1L_2\halfd\log(1+\sqrt{2})}\eea
This double sum is finite for $d>0$ and hence $\C F$ is analytic at $\mu=\mu_c$; we are left with
\beq \C Z(\mu,d)=(\mu-\mubar(1)d)^{\halfd-1}+{\rm regular}\label{L1}\eeq
except when $d=2m$ ($m$ integer) in which case there is the usual 
logarithmic modification.

Equation \rf{L1} should be compared with \rf{T3}.   We see that
the continuum calculation is misleading for $d>0$ and that in fact the system undergoes
a Bose condensation in which (nearly) all the tori crowd into the single state with largest
possible modulus.  This also implies that the continuum result,
$\gstr=2-\halfd$ for $d>0$, is  wrong;  the correct result is $\gstr=3-\halfd$. Similar
considerations will apply to model II.

The variation of correlation functions  of operators, for example the $X$ fields,
 as a function of their geodesic separation, $l$, provides another 
measure of the dimensionality of the space in gravitational systems \cite{ds,kawai}. Clearly in our
models we will obtain results typical of a two dimensional system ($\sim\log l$) for $d<0$ and of a one
dimensional system for $d>0$.  However it is not immediately obvious how the correlation function
will behave at $d=0$ where all tori contribute equally; letting the separation $l$ lie along the real
direction (and ensuring that it is the shortest distance) we find the $X$ two point function
\beq \C G(\mu,l)\simeq -\half \int_{2l}^\infty du\, e^{-\mu u}\int_{{1\over u}}^{{u\over 4 l^2}}
{dt\over t}\,\log\Gamma_{12}(\tau=it,\omega_1=\sqrt{{u\over t}},z_{12}=l)\eeq
where $\Gamma_{12}$ is given in \cite{ItZu}; replacing the Jacobi function $\vartheta_1$
by its small argument approximation 
 we obtain
\bea \C G(\mu,l)&\simeq& - \half\int_{2l}^\infty du\, e^{-\mu u}\int_{{1\over u}}^{{u\over 4 l^2}}
{dt\over t}\,\log{l^2}\nn\\
&=&-{1\over\mu}\left\{-2\log\mu\log l-2(\log l)^2+O(\log l)\right\}\eea
The leading $l$ dependent contribution at the critical point $\mu\to0$ is $\log l$
but the presence of sub-leading contributions $(\log l)^2$ which are more singular
in $l$ indicates that the dimensionality of the system is unstable.

The dimensional collapse which occurs in models I and II at $d=0$ may be analogous to
the geometry change which is believed to occur at $c=1$ in the full model of 
conformal matter coupled to quantum gravity. The difference betweenthe change 
occuring at zero families  of scalar fields or one is trivial; it simply arises because the 
number of graphs of a given type in the full model is exponential in $N$ which moves
the competing singularities around.  On the other hand the free energy gap which 
generates the Bose condensation discussed above is absent in the full model
(which, in this respect alone, is presumably more like the integral/continuum 
version of I and II).   It appears from numerical work \cite{ADJT} that the geometry
change in the full model is less dramatic than in I and II; the system does not 
seem to collapse to a collection of quasi-one-dimensional objects as soon as $c$ exceeds 1.

\vspace{1 truecm}
\noindent We acknowledge valuable conversations with J.L. Cardy and I.I. Kogan.  
J.C. acknowledges  a grant from {\sc Praxis XXI}.


\begin{thebibliography}{99}
\bibitem{BP}{J.Ambj\o rn, B.Durhuus and J.Fr\"ohlich,
Nucl. Phys. B 257 [FS14](1985) 433.}
\bibitem{FD}F.David,
Nucl. Phys. B 257 [FS14](1985) 543.
\bibitem{KKM}V.A.Kazakov, I.K.Kostov and A.A.Migdal, Phys. Lett. 157B
(1985) 295.
%
\bibitem{KPZ} V.G.Knizhnik, A.M.Polyakov and A.B.Zamolodchikov, Mod. Phys.
Lett. A3 (1988) 819;\\
F. David, Mod. Phys. Lett. A3 (1988) 1651;\\
J. Distler and H. Kawai, Nucl. Phys. B321 (1989) 509.





 
\bibitem{Bre}{E.Br\'ezin and S.Hikami,
  Phys. Lett. B283 (1992) 203,
  Phys. Lett. B295 (1992) 209;
S.Hikami, Phys. Lett. B305 (1993) 327.}

%Matrix models on large graphs
\bibitem{Wex}{M.Wexler,  Nucl. Phys. B410 (1993) 377, Mod. Phys. Lett. A8 (1993) 2703.}

%Matter fields with c>1 coupled to 2d gravity
\bibitem{ADJT}{C.F.Baillie and D.A.Johnston,
  Mod. Phys. Lett. A7 (1992) 1519;\\J.Ambj{\o}rn, B.Durhuus,  T.J\'onsson, G. Thorleifsson,
  Nucl. Phys. B398 (1993) 568.}
\bibitem{mghjfw} M.G. Harris and J.F. Wheater, Nucl. Phys. B427 (1994) 111.
\bibitem{mghamb} M.G. Harris and J.Ambj{\o}rn, Niels Bohr Institute preprint NBI-HE-96-04.
\bibitem{RR}V.A. Kazakov, M. Staudacher and T. Wynter, Nucl. Phys. B471 (1996) 309.
\bibitem{kazakov}V.A. Kazakov, M. Staudacher and T. Wynter, \'Ecole Normale preprints
LPTENS-95/9,24, hep-th/9502132, 9506174, Commun. Math. Phys. to be published.


\bibitem{kk}{D.V.Boulatov and V.A.Kazakov, Phys. Lett. B186
(1987) 379.}
\bibitem{ItZu}{C.Itzykson and J.B.Zuber, Nucl. Phys. B275[FS17] (1986) 580.}

\bibitem{ds} {J.Ambj{\o}rn, J.Jurkiewicz and Y.Watabiki, Niels Bohr Institute preprint
NBI-HE-95-22.}
\bibitem{kawai}{H.Aoki, H.Kawai, J.Nishimura and A. Tsuchiya, KEK preprint KEK-TH-454.}
\end{thebibliography}
\end{document}